\documentstyle[aps,twocolumn,definitions,10pt,psfig]{revtex}
\def\beqra{\begin{eqnarray}} 
\def\eeqra{\end{eqnarray}}
\def\beqran{\begin{eqnarray*}} 
\def\eeqran{\end{eqnarray*}}
\def\beq{\begin{equation}}      
\def\eeq{\end{equation}}

\newcommand\p{\partial}

\def\pb{|{\bf p}|}
\def\i{i}
\def\d{d}

\def\half{\mbox{\small $\frac{1}{2}$}}

\def\Red{}
\def\Black{}
\def\Blue{}

\begin{document}
\twocolumn[\hsize\textwidth\columnwidth\hsize\csname
@twocolumnfalse\endcsname

\title{\Red Wilsonian flow equation and thermal field theory\Black$^\dagger$ }
\author{
\hfill 
Daniel F. Litim$^*$
\hfill\raisebox{21mm}[0mm][0mm]{\makebox[0mm][r]{ECM-UB-PF-98-23}}%
}
\address{Departament ECM,
        Facultat de $F\!\raisebox{.15ex}{\'}\!\!\!\,{\i}sica$ {\sl \&} IFAE, Universitat de Barcelona,
        Diagonal 647, E 08028 Barcelona, Spain.}
\maketitle

\begin{abstract}\noindent
{\Blue We review the use of Wilsonian renormalization group methods for quantum field theories at finite temperature. The implementations within both real and imaginary time formalism is carefully discussed. In particular, the question of gauge invariance is addressed in detail. A recently proposed real time thermal renormalization group is then used to derive the RG flows for Abelian Higgs models. Further applications are outlined.\Black}
\end{abstract}
\vskip1.5pc]

\section{Introduction}
\footnotetext{$^\dagger$Based on a talk presented 
at the $5^{th}$ International Workshop on Thermal Field Theories 
and their Applications, Regensburg, Germany, August 1998.}
\footnotetext{$^*$E-Mail: Litim@ecm.ub.es}

Most of the physically interesting questions in thermal field theory are outside the domain of validity of perturbation theory. This is true not only for static quantities like the magnetic mass or the equation of state for a quark-gluon plasma, but as well as for dynamical ones, like the plasmon damping rate close to the critical temperature.  The reason for this break-down are IR divergences, which are difficult to control perturbatively.

The use of reliable resummation procedures seems therefore mandatory. The Wilsonian or Exact Renormalization Group is precisely a tool that allows for a systematic resummation beyond perturbation theory. 

All current implementations have in common, that they use the ``known'' physics in the UV as the starting point (see fig.~1). This region  -the upper right corner-  corresponds to the bare classical action of the $T=0$ theory. The goal is to find the corresponding effective action of the soft modes at non-vanishing temperature.  This region -the lower left corner- describes the IR limit. The resummation problem is thus the question about how these two regions are related.
  
For vanishing temperature, the flows towards the IR only integrate-out quantum fluctuations, and are depicted by the flows along the vertical boundaries (for the $4d$ and $3d$ limits, respectively). For $T\neq 0$, one might distinguish essentially three scenarios. 

1. The {\it dimensional reduction} approach aims at relating the $4d$, $T=0$ parameters to those of an effective $3d$ theory at $T=0$. This reduces the problem to a purely $3d$ one, that is, to the problem of integrating out only quantum fluctuations in $3d$. The temperature enters via the initial parameters of the effective $3d$ theory.

2. Flow equations in the {\it imaginary time} formalism are used to directly relate the $4d$ couplings in the UV region at $T=0$ with the renormalized ones at $T\neq 0$. Both quantum and thermal fluctuations are integrated out, as the imaginary time formalism does not distinguish between them. The Euclidean flow equation in $4d$ can directly be used, with the standard prescriptions of the Matsubara formalism. This scenario corresponds to the flow along the diagonal in fig.~1. 
\begin{figure}[ht]
\psfig{file=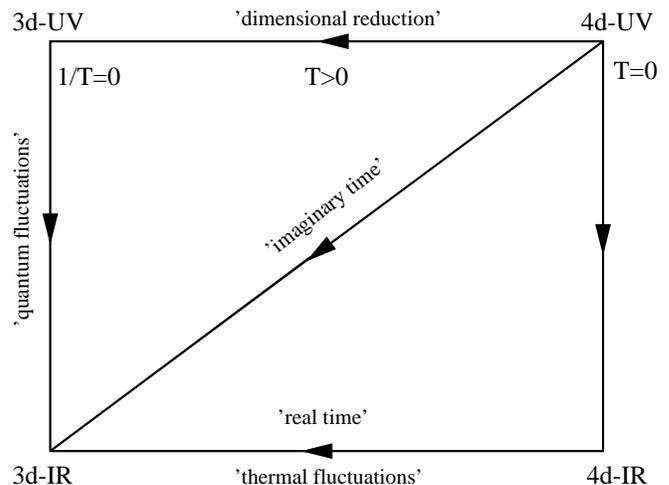,width=\hsize}
\vskip.3cm
\begin{center}
\begin{minipage}{\hsize}
\caption{\small The qualitative difference of renormalization group flows for theories at finite temperature. The arrows indicate the flow towards the infrared.}
\end{minipage} 
\end{center}
\end{figure}
3. Flow equations in the {\it real time} formalism are used to relate the $4d$ {renormalized} couplings at $T=0$ with the renormalized ones at $T\neq 0$. A prerequisite of this approach is of course the knowledge of the renormalized $4d$ couplings in the first place. Contrary to the imaginary time approach, this one allows the investigation of non-static properties. The flow equation only integrates-out thermal fluctuations, that is, it describes how modes with momenta around $k$ come into thermal equilibrium at temperature $T$. This RG flow corresponds to the flow along the base line in fig.~1.

The paper is organized as follows: The sections II and III aim at giving an introduction to the key aspects of Wilsonian RGs. Sect.~II reviews the  Euclidean formalism, and in particular the implementation at non-vanishing temperature and the inclusion of gauge fields. Sect.~III is reserved for a recently proposed  real time implementation. Sect.~IV presents an application of the latter to the U(1)-Higgs model, while sect.~V contains the discussion and an outlook. 

\section{Wilsonian flow in Euclidean time}
In this section we will outline the Exact Renormalization Group approach to quantum field theories in Euclidean space time \cite{Wilson,ERG,AverageAction,Flows}. In particular, we discuss the implementation for gauge theories \cite{Abelsch,ReuterWetterich,Ellwanger,Axial,Marchesini}, and the application to thermal field theories \cite{averageT,StevensConnor,LiaoStrickland,TetradisT,FreireLitim} in the imaginary time formalism.

\subsection{Coarse graining}
The main problem of perturbative methods for field theories at vanishing or finite temperature can be linked to the problematic IR behaviour of massless modes in less than four dimensions. It is therefore mandatory to find a  regularization for them.  Let us consider the case of a bosonic field $\phi$. A particularly simple way of curing the possible IR singular behaviour of its perturbative propagator $P_\phi$ consists in replacing it by a  cut-off propagator
\beq\label{euclidcoarse}
P_\phi\to P_\phi\ \Theta_k\left({p^2}/{k^2}\right) \ .
\eeq
Here, we introduced a function $\Theta_k$ which depends on a yet unspecified additional momentum scale $k$. $\Theta_k$ is meant to be a (smeared-out) Heavyside step function: for large momenta $p\gg k$ it goes to one (no regularization is needed), while it vanishes (at least with $p^2$) for $p\ll k$. For any $k>0$, the above propagator remains IR finite and could safely be used within loop integrals. Finally, however, we are interested in the $k\to 0$ limit in which the regulator is removed. Thus we need to describe the $k$-dependence of the theory, which brings into life the Exact Renormalization Group. Originally, it has been interpreted as a coarse-graining procedure for the fields, averaging them over some volume $k^{-d}$. In this light, the IR limit corresponds to averaging fields over bigger and bigger volumes, {\it i.e.}~to the limit ${k\to 0}$.

\subsection{The exact renormalization group}
A path integral implementation of these ideas goes back to \cite{Wilson,ERG}, where a regulator is used in order to distinguish between hard $(p^2>k^2)$ and soft modes $(p^2<k^2)$. A slightly different point of view has been taken in \cite{AverageAction} (see also \cite{Flows}), where a smooth cut-off has been employed. Following these lines, one obtains an effective theory for the soft modes only. The starting point is the functional
\beq \label{Schwingerk}
\exp W_k[J]=\int{\cal D}\phi \exp\left(-S_k[\phi] + \int \0{d^d p}{(2\pi)^d} 
J(-p) \phi(p) \right)
\eeq
Here, $\phi$ stands for all possible fields in the theory which we shall restrict to be bosons, for simplicity. (The extension to fermions is straightforward.) $J$ are the corresponding sources, and $S_k=S+\Delta_k S$ contains the (gauge-fixed) classical action $S[\phi]$ and a quadratic term $\Delta_k S[\phi]$, given by
\beq 
\Delta_k S[\phi] = \012 \int \0{d^d p}{(2\pi)^d}\ \phi^*(-p)\ R_k(p)\ \phi(p) 
. \label{Rk}
\eeq 
It introduces a coarse-graining via the operator $R_k(p)$. Performing a Legendre transformation leads to the coarse-grained effective action $\Ga_k[\phi]$,
\beq
\Ga_k[\phi]=-W_k[J]-\Delta_kS[\phi]+\Tr\ J \phi,\ \phi=\0{\de W_k}{\de J},
\eeq
where the trace $\Tr$ sums over momenta and indices. It is straightforward to obtain the flow equation for $\Ga_k$ w.r.t.~$t=\ln k/\Lambda$ (with $\Lambda$ some UV scale). The only explicit $k$-dependence in \eq{Schwingerk} stems from the regulator $R_k$, thus 
\beq
\label{flowE} \partial_t\Gamma_k=\frac{1}{2}{\rm Tr}
\left\{G_k[\phi]\ \frac{\partial
    R_k}{\partial t}\right\} 
\eeq 
with
\beq
G_k[\phi]=\left(\frac{\delta^2\Gamma_k}{\delta  \phi\delta \phi^*}+R_k\right)^{-1} 
\eeq
denoting the full (field-dependent, regularized) propagator. For the time being, the regulator function is kept arbitrary. It can be chosen in such a way (see below), that
\bea
\lim_{k\to \infty}\Ga_k&=&S\label{initial}\\ 
\lim_{k\to 0}\Ga_k&=&\Ga\ .\label{final}
\eea
Therefore the flow equation connects the (gauge-fixed) classical action $S$ with the full quantum effective action $\Ga$. Solving the path integral \eq{Schwingerk} (for $k=0$) is therefore equivalent to solving \eq{flowE} with the initial condition \eq{initial} given at some UV scale. One might read this approach as a path integral independent definition of a quantum field theory. 

Note that the flow equation \eq{flowE} is exact: no approximations have been employed for its derivation. This means in particular that \eq{flowE} is non-perturbative -- it describes unambiguously the resummation of {\it all} (quantum and/or thermal) fluctuations. 

{\it Solving} the flow equation necessitates, however, truncations and approximations. One can easily recover the perturbative loop-expansion, or resummations thereof. However, \eq{flowE} allows for more elaborate expansion schemes which are not confined to regions of small coupling constants. Commonly used is the derivative expansion, or an expansion in powers of the fields. Deriving \eq{flowE} w.r.t.~the fields gives then flow equations for the higher order vertices, which parametrize the effective action. 

\subsection{The regulator function}
Let us be slightly more explicit about the regulator function $R_k$. We will impose the following constraints on the regulator $R_k$ such that IR finiteness, \eq{initial} and \eq{final} are ensured:
\begin{itemize}
\item[(i)] $R_k$ has a non-vanishing limit for $p^2 \to 0$, typically like a mass term $R_k\to k^2$. 
\item[(ii)] $R_k$ vanishes in the limit $k\to 0$, and for $p^2 \gg k^2$. 
\item[(iii)] For $k\to \infty$ (or $k\to \Lambda$ with $\Lambda$ being some UV scale much larger than the relevant physical scales), $R_k$ diverges like $\Lambda^2$. 
\end{itemize}
Condition (i) reflects the IR finiteness of the propagator at non-vanishing $k$ even for vanishing momentum $p$ (which is why the regulator has been introduced in the first place). Condition (ii) ensures that any dependence on $R_k$ drops out for $k\to 0$ (that is to say that $\Gamma_{k\to 0}$ reduces to the full quantum effective action $\Gamma$), and that large momenta modes are suppressed, ({\it i.e.}~integrated-out) . From condition (iii) we conclude that the saddle point approximation to (\ref{Schwingerk}) becomes exact for $k\to\La$,  and $\Gamma_{k\to\Lambda}$ reduces to the (gauge-fixed) classical action $S$. The regulator function is related to $\Theta_k$ in \eq{euclidcoarse} as
\beq\label{theta}
\Theta_k\left(\0{p^2}{k^2}\right)=1-\0{R_k(p^2)}{p^2+R_k(p^2)}.
\eeq
One easily verifies that any $R_k$ with the properties (i)-(iii) yields a (smeared-out) Heavyside step function, when inserted in \eq{theta}. We also conclude from \eq{theta} that the operator $\partial_t R_k$ is a (coarse-grained) $\delta$-function. This is consistent with the flow equation \eq{flowE}. At any fixed scale $k$, only loop-momenta $p^2$ around $k^2$ can contribute to the change of $\Ga_k$. All other momenta are suppressed because of  $\partial_t R_k$ having support only in the vicinity of $k^2$. This is the essence of a Wilsonian philosophy based on the integration over infinitesimal momentum shells.
 
Typical classes of smooth regulators used in the literature are exponential ones with
\beq\label{expreg}
R_k(p^2)= \0{p^2}{\exp(p^2/k^2)^n-1}
\eeq
or algebraic ones with
\beq
R_k(p^2)= p^2\left(\0{k^2}{p^2}\right)^n\ .
\eeq
For $n=1$, both classes have a mass-like limit for small momenta. The limit $n\to\infty$ corresponds to the sharp cut-off limit \cite{Scalar}. 

Two comments are in order. The first one concerns the case of a mass-like regulator $R_k=k^2$. This regulator is somewhat special. First of all, it is {\it independent} of momenta. The operator  $\partial_t R_k$ in \eq{flowE} is -for this particular regulator- neither peaked nor sufficiently suppressed for large momenta. Thus, {\it all} momenta $p$ do contribute to $\Gamma_k$ at any $k$ ({\it i.e.}~the second part of (ii) is not fulfilled). Typically it is observed that the convergence properties of approximate solutions to the flow equation are worse for a mass-like regulator then for exponential ones. Furthermore, the unsuppressed large momenta modes introduce additional UV divergences. Although analytical computations simplify tremendously with this regulator, it has to be taken with care. 

The second comment concerns the (in-)dependence of physical quantities on the shape of the regulator function. Obviously, physical observables should not depend on the particular regulator chosen \cite{Litim97a}, and will not depend on $R_k$, if the flow equation is integrated down to $k=0$. This is an immediate consequence of \eq{final}. However, any computation has to resort to some approximation (typically only a finite number of operators are considered for an expansion of $\Ga_k$). Thus, approximations can introduce a {\it spurious} scheme dependence. Any approximation scheme, that yields large quantitative or even qualitative corrections is not acceptable and has to be discarded. Therefore it is mandatory to compute the scheme dependence within a given approximation. At the same time, this is an efficient way to check the viability of a given Ansatz \cite{Litim97a}.

\subsection{Gauge invariant Green functions}
What can be done for gauge theories? The obvious problem is that the regulator term -being quadratic in the fields- is not gauge invariant, which raised some criticism about the present approach. The question arises as to which extent gauge theories can be handled, although the regulator term \eq{Rk} seems not to be compatible with gauge invariance. This problem has been considered using the background field method \cite{Abelsch,ReuterWetterich}, modified Slavnov-Taylor or Ward Identities \cite{Ellwanger,Axial} or (perturbative) fine tuning conditions \cite{Marchesini}.

We will follow \cite{Axial} to illustrate the problem, and to show that gauge invariance of physical Green functions can indeed be maintained. Let us consider the example of an SU($N$) gauge theory in an axial gauge, with $\phi=A^a_\mu$ and the gauge fixing
\beq
\label{gf}
S_{\rm gf}[A]=\01{2}\Tr \ n_\mu A^a_\mu\ \01{\xi n^2}\ n_\nu A^a_\nu.  
\eeq
Here, $\xi$ is the gauge fixing parameter (chosen to be momentum independent) and $n_\mu$ is a fixed Lorentz vector. Axial gauges have the nice property that the ghost sector decouples. The problem of Gribov copies is therefore absent, and the number of Feynman diagrams is significantly reduced. Furthermore, the spurious singularities as encountered within a perturbative approach have been shown to be absent \cite{Axial}. Finally, the axial vector $n_\mu$, which appears in the gauge fixing condition, has a natural explanation within thermal field theories as the thermal bath singles-out a rest frame characterized by a Lorentz vector.

Let us now perform an infinitesimal gauge transformation. This leaves the measure in \eq{Schwingerk} invariant and leads to  a functional identity, the so-called {\it modified} Ward Identity (mWI) ${\cal W}_k[A]=0$, with
\bea \di {\cal
  W}_k^a[A]&=& \di D_\mu^{ab}(x)\frac{\delta
  \Gamma_k[A]}{\delta A^b_\mu(x)} \di-\frac{1}{n^2\xi} 
n_\mu\partial^x_\mu\ n_\nu A^a_\nu (x) \nonumber \\  && \di -g \int d^dy f^{abc}R^{cd}_{k,\mu\nu}
(x,y) G_{k,\nu\mu}^{db}(y,x).  
\label{mWI}
\eea 
and $D^{ab}_\mu=\de^{ab}\partial_\mu+ g f^{acb} A_\mu^c$. The mWI contains all terms of the standard Ward Identity (WI), {\it i.e.}~the first line of \eq{mWI}. Additionally it contains also a term proportional to $R_k$, which is a remnant of the coarse-graining. We observe that this term vanishes for any regulator in the limits $k\to\infty$ and $k\to 0$. In particular, it vanishes identically for a mass-like regulator $R_k=k^2$, thus reducing the mWI to the usual WI in these cases. The flow of ${\cal W}_k[A]$ is obtained 
from \eq{flowE} and \eq{mWI}:
\beq \label{compatible}
\partial_t {\cal
  W}^a_k = -\frac{1}{2}{\rm Tr}\left( G_k \frac{\partial
    R_k}{\partial t} G_k \frac{\delta}{\delta
    A}\times \frac{\delta}{\delta A}\right){\cal W}^a_k \ .
\eeq
The flow \eq{compatible} has a fixed point ${\cal W}_k[A]=0$. Thus, if $\Ga_k[A]$ solves \eq{mWI} at some scale $k_0$ and evolves subsequently according to \eq{flowE}, then $\Ga_k[A], k<k_0$, fulfills as well ${\cal W}_k[A]=0$. In particular, $\Ga_{k=0}[A]$ will obey the usual WI, which establishes gauge invariance for the {\it physical} Green functions.

This approach has been used to prove that the 1-loop $\beta$-function of SU($N$) gauge theory coupled to $N_f$ fermions is indeed universal, independent of the choice for $R_k$ or $\xi$ (in $d=4$) \cite{Axial,1-loop}. 

It is worth stressing that the mWI plays a double role when it comes to {\it approximate} solutions, {\it i.e.} for a truncation of $\Ga_k[A]$. First of all, the mWI can be implemented even in these cases. Perturbatively, this is well known, and sometimes denoted as the Quantum Action Principle. A general procedure that allows to respect the mWI for numerical implementations even {\it beyond} perturbation theory can also be given (for the details, see \cite{thermalRG}). At the same time, the mWI allows the control of the domain of validity for a given truncation \cite{thermalRG}. This error control is quite welcome also on a computational level as it avoids to go to the next order in the chosen expansion.

\subsection{Imaginary time formalism}
It is straightforward to upgrade the above approach to the case of non-vanishing temperature within the imaginary time formalism \cite{averageT}. The flow equation contains a loop integral over some momentum dependent functional. Therefore, the only changes concern the Tr in the flow equation, which becomes a sum over Matsubara frequencies, and the 0-component of the loop momenta, which is discretized
\bea
\int\0{d^dp}{(2\pi)^d} \to T\sum_n \int\0{d^{d-1}p}{(2\pi)^{d-1}},\quad p_0\to 2\pi n T \ .
\eea
Note, however, that the flow equation now connects the UV parameters of the $4d$ theory at $T=0$ with the IR ones at $T\neq 0$. Thus, both {\it quantum} and {\it thermal} fluctuations do contribute to the flow equation. This procedure corresponds to integrating along the diagonal as depicted in fig.~1. It has been applied to phase transitions in scalar \cite{averageT,StevensConnor,LiaoStrickland}  and gauge field theories \cite{TetradisT,FreireLitim}.

At this point it is interesting to note the similarity between exact flow equations and the approach advocated in \cite{Pressure} to compute the non-perturbative pressure. Indeed, the method of \cite{Pressure} can be seen as a flow equation with a mass-like regulator $R_k=k^2$. But as we commented earlier, a mass term regulates only {\it marginally}. In order to avoid the additional UV problem for large momenta fluctuations, one should consider instead {\it differences}, like $P[T]-P[0]$. This makes these divergences to cancel out and yields a well-defined flow equation. Even more interesting is the extension to gauge theories which has not been studied yet. For a mass-like regulator the second line in \eq{mWI} vanishes identically. This corresponds to the statement that gauge invariance can be maintained for any $k$ (in this particular case), and is a special feature of the axial gauge fixing used. The above leads to the conclusion that the generalization of \cite{Pressure} to gauge theories is most conveniently done within axial gauges \cite{Axial}. A more detailed account is given in \cite{thermalRG}.

\section{Wilsonian flow in real time}
The philosophy of the previous section is most appropriate for static situations, that is to say for equilibrium physics, and can be used to compute physical quantities at first order phase transitions ({\it i.e.}~free energies, surface tension, latent heat), or at second order ones ({\it i.e.}~critical exponents, equations of state, amplitude ratios). Yet a number of interesting physical problems are related to non-static properties of quantum field theories, and the question raises whether this approach can be extended to space-time with Minkowskian signature. 

Recently, a strategy has been proposed for integrating-out the temperature fluctuations within a real-time formulation of thermal field theory \cite{TRG1}. The key idea is to introduce a {\it thermal} cut-off for the on-shell degrees of freedom. This philosophy has several advantages. By construction, it allows to study precisely the effects of thermal fluctuations only, and it is not restricted to static quantities \cite{Pietroni}. As thermal fluctuations are on-shell, it is straightforward to guarantee a gauge invariant implementation of this coarse-graining even for intermediate coarse-graining scale $k$ \cite{TRG2}. However, no statements can be made regarding quantum fluctuations. They have to be included from the onset in the initial condition.
 
\subsection{Real time formalism}
At finite temperature the fields $\phi$ are defined on a contour $C$ in the complex time plane \cite{TFT,Kapusta,LeBellac}. In the real time formulation this contour consists of a forward branch (the real axis in the complex time plane) and a backward branch (parallel to the forward branch, at arbitrary distance $\sigma<1/T$). It is convenient to introduce two separate set of fields $\phi_1$ (the original fields) and $\phi_2$ (the so-called thermal ghosts), living on the two branches. Thus there is a doubling of the degrees of freedom and the propagators become $2\times 2$ matrices, determined by the boundary conditions.
Let us consider the case of a scalar field of mass $m$. Its free propagator $D^{\phi}$ at $T=0$ is
\beq
D^{\phi}(p)=\frac{1}{p^2-m^2+\i\varepsilon}
\eeq
The corresponding tree level real time propagator in momentum space  is 
\beq
D^{\phi}_T=Q[D^{\phi}(p)] + \Delta^{\phi}(p)\, N_T(|p_0|)
\; B.\label{thermalprop}
\eeq
We introduced $\Delta^{\phi}(p)=D^{\phi}(p) - \left(D^{\phi}(p)\right)^*$ and the matrices $B_{ij}=1, (i,j=1,2)$ and
\bea
Q[D^{\phi}] &=& 
\left(
   \begin{array}{cc}
   D^{\phi} &  \Delta^{\phi}\theta(-p_0) \\   &\\
   \Delta^{\phi}\theta(p_0) & - \left(D^{\phi}\right)^*
   \end{array}
\right) 
\label{nonthermal}
\eea
The function $N_T$ denotes some thermal distribution function, which, at thermal equilibrium, would be the Bose-Einstein distribution. It contains basically all the thermal information. In the $\varepsilon\rightarrow 0$ limit, we observe 
\beq
\Delta^{\phi}(p) \longrightarrow -2\i \pi \de(p^2-m^2)
\eeq
and conclude that the thermal part of the tree level propagator involves on-shell degrees of freedom only.

\subsection{Thermal coarse graining}
The question is how the thermal propagator could be modified in order to implement a thermal coarse graining. The proposal of \cite{TRG1} consists in a scale dependent modification of the thermal distribution function,
\beq\label{thermalcoarse}
N_T\to N_{T,k}=N_T \, \Theta_k(\pb/k)
\eeq 
The corresponding tree level cut off thermal propagator $D_{T,k}$ obtains from \eq{thermalprop} through the replacement \eq{thermalcoarse}, {i.e.}
\beq
D^{\phi}_{T,k}=Q[D^{\phi}(p)] + \Delta^{\phi}(p)\, N_{T,k}(|p_0|,\pb)
\; B.\label{thermalprop-k}
\eeq
Eq.~\eq{thermalcoarse} may be interpreted as the thermal analogue of \eq{euclidcoarse}. The thermal distribution $N_T$ is switched on mode by mode through the $\Theta_k$-function, whereas in  \eq{euclidcoarse}, the propagation of longer wave length modes is switched on.
The modes with $|{\bf p}|\gg k$ will be in thermal equilibrium at temperature $T$, while those with $k\gg |{\bf p}|$ remain in equilibrium at the temperature $T=0$. We shall use in the sequel a sharp cut-off 
\beq
\Theta_k(\pb)=\theta(|{\bf p}|-k).
\eeq
Alternatively, an exponentially smooth cut-off is given by
\beq\label{smooth}
\Theta_k=1-\0{e^{|p_0|/T}}{1+e^{\pb/k}(e^{|p_0|/T}-1)}\ .
\eeq  
Note that \eq{smooth} yields a modified Bose-Einstein distribution very similar to the one proposed  by Nair -although within a different prospective- in \cite{Nair}. 

\subsection{The thermal renormalization group}
A path integral formulation of these ideas has been given, following \cite{TFT}, in full analogy to the Euclidean case \cite{TRG1}. Let us first introduce sources $J_i\ (i=1,2)$  for the fields $\phi_i$ in order to obtain the path integral representation of the generating functional of real-time cutoff Green functions as
\bea 
Z_k[J] &=& \int {\cal D}\phi_1 {\cal D}\phi_2 \exp \i \big( \half 
\Tr\,\phi_i  \left(D_{T,k}^{-1}\right)_{ij}  \phi_j 
\nonumber \\ && \quad\quad\quad +  S_{\rm int}[\phi] + \Tr\,J_i\phi_i\big)\,.
\label{path}
\eea
The trace corresponds to the sum over all fields and indices (only the thermal one have been given explicitly), and momentum integration. $S_{\rm int}[\phi] $ is the bare interaction action
\beq
S_{\rm int}[\phi]=S_{\rm int}[\phi_1]-S^*_{\rm int}[\phi_2].
\eeq
The flow equation for $Z_k[J]$ can be derived easily from \eq{path} and reads, using $t=\ln (k/\Lambda)$,
\beq
\p_t Z_k[J] 
=-\0{\i}{2} {\rm Tr}\left[\, \0{\de}{\de J_i}  \p_t \left(D_{T,k}^{-1}\right)_{ij}
\0{\de}{\de J_j}  Z_k[J]\right] \,.
\label{evz}
\eeq
We define as usual the cutoff effective action $\Ga_k$ as the Legendre transform of the generating functional of the connected Green functions, using $W_k[J]= \i \ln Z_k[J]$, as 
\beq
\Ga_k[\phi] = \Tr\,J_i\phi_i - \s012\,\phi_i\left( D_{T,k}^{-1}\right)_{ij} \phi_j -W_k[J] 
\label{action}
\eeq
with
\beq
\phi_i =\0{\de W_k[J]}{\de J_i}
\eeq
where we have isolated the free part of the cutoff effective action and used for the classical fields the same notation as for the quantum fields.

The flow equation for the cutoff effective action $\Ga_k[\phi]$ follows as
\bea\label{flow}
\p_t \Ga_k[\Phi] &=& \0{\i}{2}  \Tr \left[\p_t  D_{T,k}^{-1}\left(D_{T,k}^{-1}  
+\0{\de^2  \Ga_k[\phi] }{\de\phi \:\de\phi} 
\right)^{-1} \right] 
\eea
(thermal indices have been suppressed now). This flow equation is the thermal analogue of \eq{flowE}.
Given that the initial condition for $\Ga_\La[\phi]$ at $\La\gg T$ is the full renormalized theory at zero temperature, the above flow equation describes the effect of the inclusion of thermal fluctuations at a momentum scale around $\pb = k$. At any fixed scale $k$, $\Ga_k$ describes a system in which only the high frequency modes $|{\bf p}| >k$ are in thermal equilibrium, while the low frequency modes $|{\bf p}| <k$ do not feel the thermal bath and behave like zero temperature modes.

The flow equations for all the possible vertices are obtained by expanding \eq{flow} in powers of the fields. For their derivation it is helpful to note that the cutoff effective action has a discrete $Z_2$ symmetry \cite{NS}
\beq
\Ga_k[\phi_1,\phi_2]=-\Ga^*_k[\phi_2^*,\phi_1^*]\ ,
\eeq
which relates different vertices to each other.

\section{Application: U(1) Higgs theory}
In this section we will apply the real-time thermal RG to an Abelian Higgs model. A derivation of the flow for the effective potential is given. A more elaborate presentation will be given elsewhere \cite{realAHM}.

The U(1) Higgs model is quite appealing as a testing ground for the feasibility of this approach. First of all, it is an important model for cosmological phase transitions. Furthermore, the starting point for the use of \eq{flow}, that is the renormalized action at vanishing temperature, is well known \cite{abelianhiggs4d} and can be computed with high accuracy. This model has several different mass scales, which we expect to thermalize and decouple at different coarse-graining scales. Finally, in the limit $T\to \infty$ we expect to find flow equations for the purely $3d$ Abelian Higgs model. This might shed new light on the superconducting phase transition in $3d$.

\subsection{The Euclidean effective potential}
The field content of this model is given by $N$ complex scalar fields, an Abelian gauge field, ghost fields, and their thermal partners.  We will use the Landau gauge throughout. For the time being, we shall be interested in the phase transition at finite temperature, whose details are encoded  in the coarse grained effective potential. Therefore, we have to relate the Euclidean effective potential $V_k(\bar\phi)$ to $V_k[\phi]$. Evaluating the cutoff effective action for constant field
s results in 
\beq
\Ga_k[\phi={\rm const}]=-V_k[\phi]\int d^4 x \,.
\eeq
We shall evaluate it for a field configuration $\phi=\hat\phi$ which gives a non-vanishing v.e.v.~$\bar\phi$ only to the real part of one component of $\phi_1$. All other fields are then set equal to zero. In \cite{NS} it was shown, that the Euclidean potential (for one real scalar field) is given by
\beq\label{euclidpot}
\0{\p V_k(\bar\phi)}{\p\bar\phi}=
\left. \0{\p V_k[\phi]}{\p\phi_1}
\right|_{\phi=\hat\phi}
\,,
\eeq
This relation can be generalized to an arbitrary number of fields \cite{realAHM}. Apart from an irrelevant constant, the effective potential is \cite{NS}
\beq
V_k(\bar\phi) =  \0{1}{2} m^2  \bar\phi^2 -\int^{\bar\phi} {\d} \phi \:\Ga_k^{(1)} ({\phi})  
\,,\label{effpot}
\eeq
where
\beq
\Ga_k^{(1)} (\bar\phi) \equiv \left.\0{\de\Ga_k[\phi]}{\de\phi_1}\right|_{\phi=\hat\phi}
\label{tadev}
\eeq
denotes the tadpole. Thus the flow equation for the Euclidean potential is 
deduced from the one for the tadpole, using \eq{flow}. 

\subsection{Approximate flow equations}
In order to obtain a closed set of flow equations, we have to employ some approximations regarding higher order vertices. We shall employ the leading order approximation in a derivative expansion. This implies that the wave function renormalization of the scalar field is neglected. Furthermore, we shall assume that the action remains at most quadratic in the Abelian gauge field. The (possible) imaginary parts of the scalar and photon self energies are neglected.

These approximations allow a closed set of flow equations for the functions $V_k(\rb)$ and $U_k(\rb)$, where $\rb=\bar\phi\bar\phi^*$, and $U_k(\rb)$ is the (field dependent) coefficient in front of the $A^2$ operator in the action. The flow for $U_k$ is related to the longitudinal part of the photon self energy at vanishing momenta. For the time being, we discard the distinction between the temporal and the spatial gauge field component. (The more general Ansatz  $U_k A^2\to U_{1,k} A_0^2+U_{2,k} A_i^2$ is able to correctly describe the decoupling Debye mode.) It is useful to introduce the following shorthand notations  
\bea
\Omega_1(\rb)&=&\sqrt{k^2+V_k'(\rb)+2 \rb V_k''(\rb)}\\
\Omega_2(\rb)&=&\sqrt{k^2+V_k'(\rb)}\\
\Omega_3(\rb)&=&\sqrt{k^2+U_k(\rb)}
\eea 
in terms of which the flow for $V_k(\rb)$ reads 
\bea
\0{\partial_t V_k(\rb)}{Tk^3} = &-&\0{1}{2 \pi^2} \ln\left[1-\exp\left(-\Omega_1/T\right)\right]\,
\theta(\Omega^2_1) \nonumber \\
&-&
\0{2N-1}{2 \pi^2} \ln \left[1-\exp\left(-\Omega_2/T\right)\right]\,
\theta(\Omega^2_2) \nonumber \\
&-& \0{3}{2 \pi^2} \ln \left[1-\exp\left(-\Omega_3/T\right)\right]\,\theta(\Omega^2_3).
\eea
An analogous flow equation is obtained for the function $U_k(\rb)$. 
Now we will rescale all the dimensionful quantities with appropriate powers of $T$ and $k$ in order to obtain dimensionless flow equations for the $N$-component Abelian Higgs model. To that end, we shall introduce the following functions (using ${}'=\partial_\r$):
\bea
v(\r,t) &=& V(\rb,t)/T k^3\\
u(\r, t)&=& U(\rb,t)/k^2\\
\r &=& \rb/kT,
\eea
and $\om_i=\Omega_i/k \ (i=1,2,3)$
\bea
w_1(\r,t)&=&\sqrt{1+v'+2\r v''}\\
w_2(\r,t)&=&\sqrt{1+v'}\\
w_3(\r,t)&=&\sqrt{1+u}.
\eea
Within this notation the flow equation for $v$ takes the form
\bea
\partial_t v &=&
-3 v + \r v'\nonumber \\
&&\di -\0{\theta(w_1^2)}{2\pi^2} \ln\left[1-\exp\left(-w_1 k/T\right)\right]
\nonumber \\
&&\di -\0{\theta(w_2^2)}{2\pi^2} (2N-1) \ln\left[1-\exp\left(-
w_2 k/T\right)\right]\nonumber \\
&&\di - 3 \0{\theta(w_3^2)}{2\pi^2} \ln\left[1-\exp\left(-w_3 k/T\right)\right]
\label{v-flow}
\eea
Note that the flow still explicitly depends on the ratio $k/T$, which is just a consequence of the presence of two independent scales.

\subsection{Low and high temperature limits}
By construction, the flow equations (for $v$ and $u$) 
only integrate-out the temperature fluctuations. This is why the 
initial conditions -given for the functions $u$ and $v$ at some 
large scales $k$- already have to be the full quantum effective 
parameters of the $4d$ theory at vanishing temperature. Therefore, the flow equations have to vanish 
in the low temperature limit since no further fluctuations need to be taken 
into account. That this is actually the case is easily seen in \eq{v-flow}.  
For $T\to 0$, the exponential factors $\exp(-w k/T) \ k/T$ suppress any 
non-trivial flow.

The high temperature limit is much more interesting. For $T\to\infty$ 
we would expect to recover the purely three-dimensional running of the 
couplings. Expanding $\ln [1-\exp(w\, k/T)]=\ln w + \ln (k/T) + 
{\cal{O}}(w\ k/T)$ we obtain
\bea
\partial_t v &=&
-3 v + \r v'
\di -\0{\theta(w_1^2)}{4\pi^2} \ln [1+v'+2\r v'']
\nonumber \\
&&\di -\0{\theta(w_2^2)}{4\pi^2} (2N-1) \ln [1+v']
\di - 3 \0{\theta(w_3^2)}{4\pi^2} \ln [1+u]
\label{v3-flow}
\eea
Note that we have suppressed terms proportional to $\theta(w^2)\ln\left( k/T\right)$, as their contribution for $w^2>0$ is field-independent.

It is interesting to compare \eq{v3-flow} with the flow equation 
directly derived in $3d$ via the effective average action approach 
\cite{Abelsch}. If we specify a sharp cut-off regulator and neglect the 
scalar anomalous dimension, the corresponding
 flow equation for the Euclidean dimensionless potential 
$v^{}_{\mbox{\tiny E}}$ reads \cite{3d-abel,3d-abelN}
\bea
\partial_t v^{}_{\mbox{\tiny E}} &=&
-3 v^{}_{\mbox{\tiny E}} + \r v_{\mbox{\tiny E}}'
\di -\0{1}{4\pi^2} \ln [1+v_{\mbox{\tiny E}}'+2\r v_{\mbox{\tiny E}}'']
\nonumber \\
&&\di -\0{2N-1}{4\pi^2} \ln [1+v_{\mbox{\tiny E}}']
\di - \0{2}{4\pi^2} \ln [1+2 e^2 \r]
\label{v3E-flow}
\eea
where $e^2=\bar e^2/k$ denotes the dimensionless gauge coupling squared.

Two comments are in order. First of all, the r$\hat{\rm o}$le of the (field-dependent) mass of the gauge field  $M^2_{\mbox{\tiny E}}=2e^2\r k^2$ has been taken over by the function $u$, with $M^2=u(\r)k^2=U_k(\rb)$. With the computation of $U_k(\rb)$ we would obtain therefore the full field dependence of the Abelian charge. In \cite{3d-abel} it was argued that the field dependence of $e^2$ might play an important r$\hat{\rm o}$le close to the critical points. Note also that the numerical coefficient in front of the last term in \eq{v3-flow} differs from that in \eq{v3E-flow} ({\it i.e.}~3 instead of 2). As mentioned earlier, this is due to the Debye mode, which, in the present approximation, can not decouple properly, and therefore still contributes to the flow in the high temperature limit.

The second point concerns the $\theta$-functions, absent in \eq{v3E-flow}. They ensure that the flow \eq{v3-flow} will not run into a singularity nor develop an imaginary part. As soon as the arguments in the logarithms tend to negative values, the $\theta$-function cuts their contribution off. In \eq{v3E-flow}, this is not so obvious. While solving \eq{v3E-flow}, however, one observes that the flow does indeed avoid the singularities automatically \cite{abelianhiggs4d,3d-abel,3d-abelN}.

\section{Discussion and Outlook}
We reviewed  main features of the Wilsonian RG and argued that they represent a systematic and efficient tool for applications to thermal field theories. In particular, we emphasized that even gauge theories can be handled systematically in both the real and imaginary time formalism. The particular differences between these two approaches have been discussed, the main one being that  the imaginary time formalism is adequate for the computation of static quantities, while the real time approach allows the study of non-static quantities and non-equilibrium situations. 

The application to an Abelian Higgs model extends previous applications to an interesting toy model for cosmological phase transition. A detailed study of the first and second order phase transition is now feasible. This will also allow for an independent check of the perturbative dimensional reduction scenario, which is at the heart of recent Monte Carlo simulations. The extension to the electroweak phase transition seems to be straightforward, although more elaborate approximations have to be employed in this case. It would be particularly interesting to study the critical points of both models. In the U(1) case, one expects two critical points (describing the second order phase transition, and the end point of the first order phase transition region). This approach might even open a door to a better understanding of the superconducting phase transition, which corresponds to the large $T$ limit. In the SU(2) case one expects to find only one critical point, the end point of the line of first order phase transitions. This fixed point was recently discovered to belong to the Ising-type universality class \cite{MonteCarlo}. A field theoretical determination of the end point and the corresponding critical exponents is still missing. Up to now only Monte Carlo simulations have been able to study this parameter range. 

With these tools at hand a number of other interesting problems can now be envisaged. An open question concerns for example the thermal $\beta$-function of QCD, which has been computed by a number of groups, with remarkably different results (see \cite{oneloop-thermal} and references therein). The Wilsonian RG, and in particular the heat kernel methods used in \cite{1-loop}, can be employed even within the imaginary time formalism and should be able to resolve this point. It seems also be possible to construct a gauge invariant thermal renormalization group  within real and imaginary time  along the lines indicated earlier \cite{thermalRG}. This would be a very useful extension of \cite{Pressure} to fermions and gauge fields.

\section*{Acknowedgements}
It is a pleasure to thank U.~Heinz for organizing a very pleasant conference, F.~Freire and J.~M.~Pawlowski for an enjoyable collaboration and a critical reading of the manuscript, M. d'Attanasio for initiating the work presented in section IV, and B. Bergerhoff and M. Pietroni for discussions. Financial support from the organizers of TFT98 is gratefully acknowledged.


\begin{thebibliography}{99}
\def\BOOK#1#2#3#4{#1 {\sc #2}, #3, #4}
\def\PRA#1#2#3#4#5{ #1   Phys.~Rev.~{\bf A #3} (19#4) #5}
\def\PRB#1#2#3#4#5{#1   Phys. Rev.~{\bf B #3} (19#4) #5}
\def\PRL#1#2#3#4#5{#1   Phys. Rev.~Lett.~{\bf #3} (19#4) #5}
\def\PRC#1#2#3#4#5{#1   Phys. Rev.~{\bf C #3}  (19#4) #5}
\def\PRD#1#2#3#4#5{#1   Phys. Rev.~{\bf D #3} (19#4) #5}
\def\PRE#1#2#3#4#5{#1   Phys. Rev.~{\bf E #3} (19#4) #5}
\def\PRep#1#2#3#4#5{#1   Phys. Rep.~{\bf  #3} (19#4) #5}
\def\NPB#1#2#3#4#5{#1   Nucl. Phys.~{\bf B #3} (19#4) #5}
\def\PLB#1#2#3#4#5{#1   Phys. Lett.~{\bf B #3} (19#4) #5}
\def\ibid#1#2#3#4#5{#1   {\it ibid.~}{\bf #3} (19#4) #5}
\def\PTP#1#2#3#4#5{#1   Prog. Theor.~Phys.~{\bf B #3} (19#4) #5}
\def\SSC#1#2#3#4#5{#1   Solid State Comm.~{\bf  #3} (19#4) #5}
\def\EPL#1#2#3#4#5{#1   Europhys. Lett.~{\bf #3} (19#4) #5}
\def\JCP#1#2#3#4#5{#1   J.~Phys. (Paris) {\bf  #3} (19#4) #5}
\def\JPA#1#2#3#4#5{#1   J.~Phys. {\bf A  #3} (19#4) #5}
\def\JPB#1#2#3#4#5{#1   J.~Phys. {\bf B  #3} (19#4) #5}
\def\JPC#1#2#3#4#5{#1   J.~Phys. {\bf C  #3} (19#4) #5}
\def\ZPC#1#2#3#4#5{#1   Z.~Phys. {\bf C  #3} (19#4) #5}
\def\JETP#1#2#3#4#5{#1   Soviet Physics JETP Lett.~{\bf #3} (19#4) #5}
\def\MPLA#1#2#3#4#5{#1   Mod.~Phys. Lett.~{\bf A  #3} (19#4) #5}
\def\PA#1#2#3#4#5{#1   Physica {\bf A  #3} (19#4) #5}
\def\PS#1#2#3#4#5{#1   Physics {\bf   #3} (19#4) #5}
\def\AP#1#2#3#4#5{#1   Ann. Phys. {\bf  #3} (19#4) #5}
\def\IJMPA#1#2#3#4#5{#1   Int.~J. Mod. Phys.~ {\bf A  #3} (19#4) #5}
\def\LNC#1#2#3#4#5{#1   Lett.~Nuevo Cimento {\bf   #3} (19#4) #5}
\def\PPR#1#2#3{#1    Preprint #3}
\def\and#1#2#3{{\bf #1} (19#2) #3}

\bibitem{Wilson}K.~G.~Wilson and I.~G.~Kogut, Phys. Rep. {\bf 12} (1974) 75; F.~Wegner, A.~Houghton, Phys. Rev. {\bf A 8} (1973) 401.
\bibitem{ERG} J.~Polchinski, Nucl. Phys. {\bf B 231} (1984) 269; T.~Hurd, Commun. Math. Phys. {\bf 124} (1989) 153; G.~Keller  and C.~Kopper, M.~Salmhofer, Helv. Phys. Acta {\bf 65} (1992) 32; G.~Keller and C.~Kopper, Commun. Math. Phys. {\bf 161} (1994) 515.  
\bibitem{AverageAction}\NPB{C.~Wetterich,}{}{352}{91}{529}; \PLB{}{}{301}{93}{90}.
\bibitem{Flows}\IJMPA{T.~Morris}{}{9}{94}{2411}.
\bibitem{Abelsch}\NPB{M.~Reuter and C.~Wetterich,}{}{391}{93}{147}; \and{B 408}{93}{91}; \and{B 427}{94}{291}; F. Freire and C. Wetterich, Phys. Lett. {\bf B 380} (1996) 337.
\bibitem{ReuterWetterich}M.~Reuter and C.~Wetterich, Nucl. Phys. {\bf B 417} (1994) 181.
\bibitem{Ellwanger}\PLB{U. Ellwanger,}{}{335}{94}{364}.
\bibitem{Axial}\PLB{D.F. Litim and J.M. Pawlowski,}{}{435}{98}{181} [{\tt hep-th/9802064}]; {\tt hep-th/980920}; {\tt hep-th/980923}. 
\bibitem{Marchesini}M.~Bonini, M.~D'Attanasio and G.~Marchesini, Nucl.~Phys.~{\bf B 418} (1994) 81;  {\bf B 421} (1994) 429; 
 {\bf B 437} (1995) 163; Phys. Lett. {\bf B 346} (1995) 87. 
\bibitem{averageT}\NPB{N. Tetradis and C. Wetterich,}{}{398}{93}{659}, \IJMPA{}{}{9}{94}{4029}.
\bibitem{StevensConnor}\IJMPA{M.A. van Eijck, D. O'Connor and C.R. Stephens,}{}{10}{95}{3343}; {\tt hep-th/9406218}. 
\bibitem{LiaoStrickland}\PRD{S.-B. Liao and M. Strickland,}{}{52}{95}{3653}.
\bibitem{TetradisT}\NPB{N.~Tetradis,}{}{444}{97}{92}.
\bibitem{FreireLitim}D.~F.~Litim, {\tt hep-ph/9708401}; F.~Freire and D.F.~Litim, in preparation.
\bibitem{Scalar}\ZPC{C.~Wetterich,}{}{57}{93}{451}; \NPB{C.~Wetterich and N.~Tetradis,}{}{422}{94}{541}; \ibid{N.~Tetradis and D.F.~Litim,}{}{B 464}{96}{492}.
\bibitem{Litim97a}\PLB{D.F.~Litim,}{}{393}{97}{103}. 
\bibitem{1-loop}D.F. Litim and J.M. Pawlowski, under completion. 
\bibitem{Pressure}A. Rebhan, these proceedings [{\tt hep-ph/9808480}].
\bibitem{thermalRG}D.F. Litim and J.M. Pawlowski, {\tt hep-th/9901063}.
\bibitem{TRG1}\NPB{M.~D'Attanasio and M.~Pietroni,}{}{472}{96}{711}.
\bibitem{Pietroni}M.~Pietroni, these proceedings [{\tt hep-ph/9809390}].
\bibitem{TRG2}\NPB{M.~D'Attanasio and M.~Pietroni,}{}{498}{97}{443}.
\bibitem{TFT} P.~Landsman and Ch.~van Weert, Phys. Rep. 145 (1987) 141.
\bibitem{Kapusta}\BOOK{J.~I.~Kapusta,}{Finite-temperature field theory}{Cambridge University Press}{1989}.
\bibitem{LeBellac}\BOOK{M. Le Bellac,}{Thermal Field Theory}{Cambridge University Press}{1996}.
\bibitem{Nair}V.~P.~Nair, these proceedings [{\tt hep-th/9809086}].
\bibitem{NS} A~Niemi and G~Semenoff, Ann. of Phys. 152 (1984) 105.
\bibitem{realAHM}D.F. Litim and J.M. Pawlowski, in preparation. 
\bibitem{abelianhiggs4d}\MPLA{D.F.~Litim, C.~Wetterich and N.~Tetradis,}{}{12}{97}{2287}.
\bibitem{3d-abel}B.~Bergerhoff, F.~Freire, D.F.~Litim, S.~Lola and C.~Wetterich, Phys.~Rev.~{\bf B 53} (1996) 5734.
\bibitem{3d-abelN}\IJMPA{B.~Bergerhoff, D.F.~Litim, S.~Lola and C.~Wetterich,}{Phase Transition of $N$-Component Superconductors}{11}{96}{4273}.
\bibitem{MonteCarlo}{K. Kajantie}, M. Laine, K. Rummukainen, M. Shaposhni\-kov and M. Tsypin,  these proceedings [{\tt hep-th/9809435}].
\bibitem{oneloop-thermal}\MPLA{M.A. van Eijck, C.R. Stevens and Ch. G. van Weert,}{}{9}{94}{309}, \PRD{P. Elmfors and R. Kobes,}{}{51}{95}{774}.
\end{thebibliography}
\end{document}